# Prediction of multitasking performance post-longitudinal tDCS via EEG-based functional connectivity and machine learning methods

Akash K Rao[1][0000-0003-4025-1042], Shashank Uttrani[1][0000-0003-2601-2125], Vishnu K Menon[1][0009-0007-9449-0934], Darshil Shah[2][0009-0000-1591-6351], Arnav Bhavsar[3][0000-0003-2849-4375], Shubhajit Roy Chowdhury[3][0000-0003-1878-6657], Varun Dutt[1][0000-0002-2151-8314]

[1] Applied Cognitive Science Laboratory, Indian Institute of Technology Mandi, Himachal Pradesh, India
[2] Ashoka Centre for Social and Behavior Change, Delhi, India
[3] School of Computing and Electrical Engineering, Indian Institute of Technology Mandi, Himachal Pradesh, India
`akashrao.iitmandi@gmail.com`

**Abstract.** Predicting and understanding the changes in cognitive performance, especially after a longitudinal intervention, is a fundamental goal in neuroscience. Longitudinal brain stimulation-based interventions like transcranial direct current stimulation (tDCS) induce short-term changes in the resting membrane potential and influence cognitive processes. However, very little research has been conducted on predicting these changes in cognitive performance post-intervention. In this research, we intend to address this gap in the literature by employing different EEG-based functional connectivity analyses and machine learning algorithms to predict changes in cognitive performance in a complex multitasking task. Forty subjects were divided into experimental and active-control conditions. On Day 1, all subjects executed a multitasking task with simultaneous 32-channel EEG being acquired. From Day 2 to Day 7, subjects in the experimental condition undertook 15 minutes of 2mA anodal tDCS stimulation during task training. Subjects in the active-control condition undertook 15 minutes of sham stimulation during task training. On Day 10, all subjects again executed the multitasking task with EEG acquisition. Source-level functional connectivity metrics, namely phase lag index and directed transfer function, were extracted from the EEG data on Day 1 and Day 10. Various machine learning models were employed to predict changes in cognitive performance. Results revealed that the multi-layer perceptron and directed transfer function recorded a cross-validation training RMSE of 5.11% and a test RMSE of 4.97%. We discuss the implications of our results in developing real-time cognitive state assessors for accurately predicting cognitive performance in dynamic and complex tasks post-tDCS intervention.

# 1  Introduction

Cognitive performance can be described as the ability to effectively process and integrate information, make decisions, tackle complex problems, and efficiently use the inherent information-processing propensity of the brain [1]. The information processing capabilities become unequivocally important in complex cognitive processes like higher-order decision-making, multitasking, situational awareness, etc., where many cognitive processes are required to act in unison [1,2]. Effectively balancing several cognitive processes, such as attention, working memory, and task switching, is what multitasking entails. The propensity to distribute and shift attention between tasks is essential for successful multitasking [1]. However, the human attentional system is limited, and attempting to handle multiple stimuli simultaneously might result in attentional dispersion and poor overall performance [2].

Over the years, several behavioral interventions like neurofeedback, music, meditation, and extended reality have been used to enhance cognitive performance over time. However, non-invasive brain stimulation techniques like transcranial direct current stimulation (tDCS) have gained considerable traction recently [3]. tDCS is usually administered by passing a mild electrical current through electrodes placed on the scalp. This current has been shown to alter cortical excitability by modulating the neuron's resting membrane potential [3]. Longitudinal tDCS has been shown to alter synaptic plasticity, an essential mechanism driving neuroplasticity [3]. The two types of synaptic plasticity, namely, long-term potentiation (LTP) and long-term depression (LTD) have been linked with learning, memory, and executive functioning in healthy adults [3]. Longitudinal tDCS has been shown to induce cellular changes while inducing neuroplasticity. Brain-derived neurotrophic factor (BDNF), a protein required for neuronal survival, development, and synaptic plasticity, has been demonstrated to be enhanced by longitudinal anodal tDCS [3]. BDNF stimulates dendritic development, synaptogenesis, and neurogenesis, allowing for the formation of new neuronal connections. Furthermore, tDCS can impact the release of neurotransmitters such as glutamate and gamma-aminobutyric acid (GABA), both of which are important in synaptic development [3].

The rise in popularity of consumer-grade Electroencephalography (EEG) as a strong instrument for monitoring brain activity, paired with machine learning's computational prowess, has opened new paths for forecasting cognitive performance [4]. The combination of EEG and machine learning has offered various advantages in predicting cognitive performance. EEG directly measures brain activity with excellent temporal resolution, allowing researchers to investigate the underlying neural correlates of cognitive processes [5]. By analyzing EEG signals, researchers have analyzed numerous aspects linked with distinct cognitive functions, such as oscillatory power, event-related potentials (ERPs), or spectral coherence [5]. Machine learning methods have been found to have the advantage of detecting discreet correlations and non-linear dynamics in EEG data that standard statistical approaches may find difficult to identify. Researchers have developed methods for estimating functional connectivity measures in the source space using EEG in the past decade, revealing insights into the neurological foundation of brain networks and their relevance to cognitive processes [6]. However, although the possibility of predicting cognitive performance using functional connectivity analysis and machine learning methods carries enormous promise, it has, unfortunately, been seldom explored in the literature. In this work, we intend to address this gap in the literature by predicting cognitive performance in a multitasking paradigm post-longitudinal tDCS administration via EEG-based functional connectivity and machine learning models. The work is novel and significant as here machine learning methods have been developed on functional connectivity analysis to predict multi-task performance. In what follows, we give a brief foray into the previous research on predicting cognitive performance using EEG and machine learning methods. We then encapsulate the experiment conducted and the various functional connectivity analysis and machine learning models employed. We conclude by discussing our results and highlighting the implications of accurately predicting changes in cognitive performance post-tDCS intervention.

## 2  Background

In recent years, considerable research has been conducted on evaluating the efficacy of longitudinal tDCS administration in enhancing cognitive performance. For example, [3] evaluated the efficacy of longitudinal anodal tDCS during multitasking training. Results revealed that anodal tDCS administered over six days also enhanced cognitive performance to untrained tasks. Researchers in [7] found that three sessions of anodal tDCS produced efficient, long-lasting transfer of working memory capacity among young adults. Researchers in [8] reported that several scientists in the world had supported the efficacy of longitudinal tDCS in cognitive augmentation, and the effects have been successfully observed through neurophysiological modalities like EEG and functional-near infrared spectroscopy (fNIRs) [8].

In recent years, research work has started focusing on applying higher-order EEG metrics and machine learning algorithms in predicting cognitive processes. For instance, researchers in [4] used cross-correlation-based functional connectivity metrics to classify mental workload, eventually providing a framework for cross- and within-task workload discrimination. Researchers in [5] employed functional connectivity metrics like mutual information and phase locking value and deep learning architectures like the convolutional neural networks to classify mental workload in a classic n-back task. Researchers in [9] employed a multi-domain convolutional neural network for identifying patterns of schizophrenia from a granger causality-based functional connectivity metric [9]. Results revealed that schizophrenic patterns could be detected from functional connectivity-based metrics and convolutional neural networks with an accuracy of 91%.

To the best of our knowledge, most of the research in cognitive performance classification using EEG metrics and machine learning models has been largely centered on mental workload and working memory, with occasional forays into predicting brain disorders. However, prediction of cognitive performance in complex, multitasking paradigms using EEG-based functional connectivity metrics and machine learning methods, especially post longitudinal tDCS administration, is lacking and much need-ed in literature. In our research, we have tried to address this gap in the literature by employing two different functional connectivity metrics (directed transfer function and phase lag index) along with a cluster of machine learning algorithms to predict cognitive performance post-longitudinal tDCS administration.

## 3 Materials and Methods

### 3.1 Experiment design

Forty participants (26 males, 14 females, mean age = 23.8 years, SD = 2.11 years) at the Indian Institute of Technology Mandi, Himachal Pradesh, India participated in the study. The experiment was divided into three phases: pre-intervention (Day 1), intervention (Day 2 to Day 7), and post-intervention (Day 10). All the subjects executed the pre-intervention phase on Day 1. The participants were initially debriefed about the experiment and the objectives to be achieved. They were then screened for the possible risks during tDCS administration using the screening questionnaire suggested by [4]. The participants then gave a written, signed consent form assuring their approval for participating in the experiment. We then acquired demographic data (age, gender, level of education, dominant hand etc.) from the participants. A 32-channel EEG data acquisition system was then set up and calibrated for all subjects (see section 3.4 for more details on the EEG data acquisition and analysis). The subjects were then made to sit in front of a 27-inch monitor so that the distance between their eyes and the monitor was approximately 50cm, corresponding to a visual angle of 45º x 30º. The participants initially undertook an acclimatization session for 5 minutes, so that they could accustomed to the different components in the NASA-MATB-II task [3] and the different controls to be used in the experiment (see section 3.3 for more details on the task). The participants then executed the task with simultaneous EEG data being acquired. The participants were then equally and randomly divided into two between-subject conditions: experimental and placebo control. The participants in the experimental condition undertook 2mA of anodal tDCS intervention for 15 minutes (see section 3.2 for more details) from Day 2 to Day 7 (six sessions) during concurrent NASA-MATB-II task training. A pre- and post- tDCS intervention questionnaire was administered to evaluate any physical effects of tDCS intervention post-administration, as suggested by [7]. The participants in the active control condition undertook sham tDCS intervention for 15 minutes from Day 2 to Day 7 (six sessions) during concurrent NASA-MATB-II task training. After a forty-eight-hour cooling period, the participants executed the NASA-MATB-II task again with simultaneous EEG data acquisition on Day 10.

### 3.2 tDCS administration

We administered tDCS using the Caputron Activadose 2 tDCS device. The electrodes were 2 cm × 2 cm in size, with the anodal and cathodal electrodes having a size of 4 cm$^2$. The current had a 2mA intensity and a 0.5 mA/cm$^2$ current density. In accord-ance with the 10-20 EEG electrode placement scheme [8], the cathodal electrode was placed over the right supraorbital region and the anodal electrode was placed over the left dorsolateral prefrontal cortex (dlPFC). The electrodes were positioned using a measuring tape in accordance with the participant's usual 10-20 EEG electrode placement technique. The electrode sponges, which were used as electrodes, were placed into the electrodes after being soaked in a potassium chloride saline solution for 10 minutes. The participants in the experimental condition had their current ramped up to 2mA over the course of 10 seconds, maintained at that level for 15 minutes, and then had their current ramped down over the course of 5 seconds. The current was built up to 2mA over the course of 10 seconds for the individuals in the sham condition, maintained at that level for 30 seconds, and then ramped down over the course of 5 seconds.

### 3.3 The National Aeronautics Space Administration – Multi-Attribute Task Battery – II (NASA-MATB-II)

The NASA-MATB-II was developed by NASA to evaluate the efficacy of multitasking and decision-making performance in a complex, dynamic, task-switching environment. As pointed out in [3], NASA-MATB-II is an ideal environment for testing the effect of mental workload and task switching-induced complexity on decision-making performance [3]. The human operator of NASA-MATB-II is required to concurrently monitor and react to four separate tasks on the computer screen. System monitoring, communication, targeting, and resource management make up the task. An illustration of the NASA-MATB-II task is as shown in Figure 1.

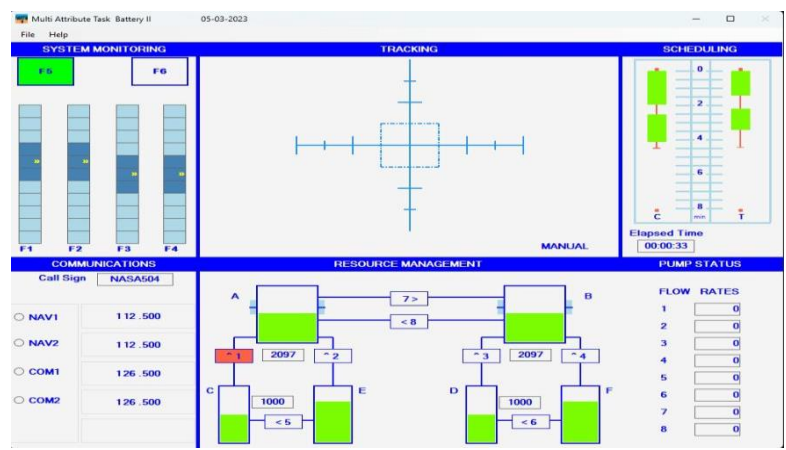

**Fig. 1.** An illustrative image of the graphical interface of the NASA-MATB-II task. The system monitoring task is at the top left of the interface, the targeting task is in the top central part of the interface, the communication task is in the bottom left of the interface, and the resource management task is in the bottom right part of the interface.

**System Monitoring.** This task, which included two subtasks (lights and dials), was in the upper-left corner of the MATB window. The lights are shown by the two rectangles at the top. The participant's goal was to keep the left light on and "display green" while the right light remained off and "displaying black." The F5 or F6 keys would reset the lights if they change from their initial state. In addition, there were four dial-filled vertical columns beneath the lights. The yellow marker inside the dials oscillated continually one position above and below the center of the dial throughout the task. If the yellow marker exceeded the indicated oscillation range, the participant would need to respond with the corresponding F1-F4 key [3].

**Communication.** This task was included in the bottom-left corner of the NASA-MATB-II window. The communications' main goal was to change the channel and frequency that had been specified in an audio cueing [3]. The participants were prompted to change a certain communication channel to a specified frequency by an auditory message. Using the up, down, left, and right arrow keys, the participants found the proper channel and set the frequency.

**Targeting.** This task was in the top middle section of the NASA-MATB-II window. A green cursor moved around the window throughout the task. The goal was to use a joystick to keep the green cursor inside of the bigger blue square.

**Resource Management.** The NASA-MATB-II window's bottom right corner included this task. The objective of resource management was to keep the level of the reservoir tanks flexible while staying within the blue-line restrictions (indicated in the tanks). This was accomplished by utilizing the 2 and 4 keys to "on" and "off" the reservoir tanks. The fluid level was continuously moving, thus it's crucial to keep in mind that the 2 and 4 keys will regularly need to be adjusted [3].

Behavioral data was also acquired from the different tasks in the NASA-MATB-II task, like reaction time (in seconds) in the system monitoring task, number of appropriate subject responses in the communication task, average root mean square error in the targeting task, and the difference between the fluid levels in the tanks in the resource management task. In this research work, we took the average root mean square error in the targeting task as the output variable predicted. The root mean square error in the targeting task was calculated by:

$$RMSE = \sqrt{\frac{\sum_{i=1}^{N} \|(x,y) - \overline{(x,y)}\|^2}{N}} \quad \ldots (1)$$

where *(x,y)* are the coordinate measurement of the green cursor in the $i^{th}$ measurement, and $\overline{(x,y)}$ is the coordinate measurement of the joystick in the $i^{th}$ measurement, and N is the number of data points in the task.

## 3.4 EEG data acquisition and analysis

The EEG data was acquired using 32 Ag/AgCl saline sensor electrodes according to the standard 10-20 system of electrode placement (Emotiv EPOC Flex and EMOTIVPRO data acquisition software, v2.34b, San Francisco, USA). The EEG data was acquired at a sampling rate of 256 Hz. The electrode impedance was kept below 10 KΩ throughout the entire session. All the main interferences were eschewed through anti-aliasing (0.1-65 Hz) with a band-pass filter and a 50 Hz notch filter. Before beginning the experiment, baseline EEG data was also acquired (for 60 seconds), where the participants were requested to relax and keep their eyes open.

The pre-processing and subsequent feature extraction from the EEG data was performed using Brainstorm [10], which is documented as a plugin in MATLAB and freely available for download under the GNU public license [10]. The raw EEG data was band-pass filtered initially from 0.1-45 Hz. We then re-referenced the band-pass filtered data to the average of the electrodes in the left and right mastoid. Eye blink artifacts were initially detected and removed by using Picard's Independent Component Analysis (ICA) algorithm [10]. We then adjusted the artifact-rejected EEG signal with respect to the baseline data acquired. We then used Brainstorm's dSPM for estimating the sources which were then parcellated into 14 different regions of interest (ROI) according to the Brodmann atlas in each hemisphere [11]. The ROIs included the primary somatosensory area, primary motor areas (anterior/posterior), pre-motor area, Broca's area (pars opercularis and pars triangularis), primary/secondary visual area, visual area in the middle temporal lobe, and the entorhinal and perirhinal cortex [10].

**Functional connectivity metrics**

*Directed transfer function based on granger causality.* The normalized directed transfer function (DTF) is given as suggested in [12] by

$$\gamma_{ab}^2(f) = \frac{|H_{ab}(f)|^2}{\sum_{m=1}^{N}|H_{ab}(f)|^2} \ldots\ldots (2)$$

$\gamma^2{}_{ab}(f)$ elucidates the ratio of the intensity of influence of $b^{th}$ channel associated cortical area on the $i^{th}$ channel associated cortical area with respect to the degree of influence of all approximated cortical signals [12]. H(f) is the transfer function of the framework and its corresponding elements $H_{ab}(f)$ represent the causal influence from the a$^{th}$ input to the b$^{th}$ output at a frequency f. We derived a 28 x 28 DTF adjacency matrix for each participant corresponding to the Brodmann ROIs extracted.

*Phase lag index.* The PLI is a model-free functional connectivity metric based on phase synchronization that assesses how phase differences are distributed among observations. It is determined by averaging the sign of the predicted phase difference for every observation [12]. It is inspired by the fact that field spread does not result in non-zero phase disparities [12]. PLI is given as:

$$PLI = |E[sign(\triangle \varphi(t))]| \ldots \ldots (3)$$

Where $\triangle t$ corresponds to the sampling period and *sign* corresponding to the signum function that repudiates the phase difference of $0|\pi|$. We derived a 28 x 28 adjacency matrix PLI adjacency matrix for each participant corresponding to the Brodmann ROIs extracted.

### 3.5 Machine learning algorithms for prediction.

**Ridge regression.** We discovered during our research that there were significantly more input variables than observations. From that point on, we employed ridge regression to minimize the parameters and decrease the complexity and potential multicollinearity. In accordance with researchers in [13], we performed ridge regression using the L2 regularization technique.

**Decision tree regressor.** A decision tree creates tree-like models for regression. It incrementally builds an associated decision tree while segmenting a dataset into progressively smaller sections [14]. The outcome is a tree containing leaf nodes and decision nodes. Two or more branches, one for each value of the characteristic under test, constitute a decision node. A choice regarding the numerical decision is represented by a leaf node [14].

**Random forest regressor.** For problems involving classification and regression, Random Forest is frequently employed. The eventual prediction in Random Forest regression is the mean of all the individual forecasts; each tree in the forest makes a prediction in this approach. High-dimensional datasets with several attributes can be addressed by Random Forest, together with missing values and noisy data [14].

**Support vector regressor.** In support vector regression, the input data points are first translated to a higher dimensional space [14]. Using a kernel function, the closest data points to the hyperplane are found, and the margin between them and the hyperplane is then maximized. SVR can handle non-linear interactions between the features and the target variable by using a non-linear kernel function [14].

**XGBoost regressor.** The XGBoost is an efficient implementation of the Gradient Boost algorithm. It is an ensemble learning method, which means it offers a prediction based on the predictive power of multiple algorithms, resulting in a single model rendering the aggregated output from several machine learning algorithms [15].

**Multi-layer perceptron.** A fully connected class of feedforward artificial neural network (ANN) is called a multi-layer perceptron (MLP). A multi-layer perceptron comprises of one input layer with one neuron (or node) for each input, one output layer with one node for each output, and any number of hidden layers with any number of nodes on each hidden layer [16].

The different hyperparameters and the corresponding values used in different machine learning algorithms are as shown in Table 1. We used 10-fold cross validation (repeated three times) [16] to select the machine learning model's best parameters. All the machine learning models were implemented using scikit-learn in Python. The machine learning models were trained for different hyperparameters and the hyperparameters obtaining the least average RMSE during training folds was considered the most optimal. We used grid search to find the best set of hyperparameters for each machine learning model. The machine learning models were trained for the different hyperparameters and the hyperparameters obtaining the least RMSE was treated to be the most optimal. The symmetric difference between the functional connectivity metrics (DTF and PLI) on Day 1 and Day 10 was taken as the input to the machine learning models and the average root mean square tracking error in the targeting task of the NASA-MATB-II was taken as the output variable to be predicted.

**Table 1.** Different hyperparameters and the corresponding range of values used in different machine learning algorithms.

| Machine learning model | Hyperparameters varied and their values |
|---|---|
| Ridge regression | 1) Solvers - Singular Value Decomposition (SVD), Cholesky, least-squares (lsqr), Stochastic Average Gradient (SAG) <br> 2) Alpha – 0.00001 to 100 (steps of 0.01) <br> 3) Fit intercept – True or False |
| Decision tree regressor | 1) Maximum depth – 2 to 10 (steps of 1) <br> 2) Minimum samples split – 2 to10 (steps of 1) <br> 3) Minimum samples leaf – 1 to 10 (steps of 1) |
| Random forest regressor | 1) Maximum depth – 2 to 10 (steps of 1) <br> 2) Minimum samples split – 2 to10 (steps of 1) <br> 3) Minimum samples leaf – 1 to 10 (steps of 1) <br> 4) Number of estimators – 10 to 100 (steps of 10) |
| Support vector regressor | 1) C – 0.01 to 100 (steps of 0.01) <br> 2) Gamma – 0.001 to 1 (steps of 0.001) <br> 3) Kernel – Linear, Polynomial, and Radial basis function |
| XGBoost regressor | 1) Learning rate – 0.01 to 0.5 (steps of 0.01) <br> 2) Maximum depth – 2 to 10 (steps of 1) <br> 3) Subsample – 0.2 to 1.0 (steps of 0.2) |

| | 4) Column sample by tree – 0.2 to1.0 (steps of 0.2) <br> 5) Number of estimators – 10 to 150 (steps of 10) |
|---|---|
| Multi-layer perceptron | 1) Hidden layer sizes – 1 to 3 (steps of 1) <br> 2) Hidden nodes count – 10 to 1000 (steps of 10) <br> 3) Activation function – Logistic, tanh, Rectified Linear Unit <br> 4) Solver – Adaptive moments (adam), Limited-memory Broyden–Fletcher–Goldfarb–Shanno (lbfgs), Stochastic Gradient Descent (SGD) <br> 5) Alpha – 0.0001 to 0.1 (steps of 0.001) |

## 4   Results

Table 2 shows show the average train and test RMSEs from the training and test folds for different machine learning algorithms combined with phase lag index as an input feature, and the best set of hyperparameters obtained from model calibration.

**Table 2.** Train and test RMSEs for the average root mean square tracking error in the NASA-MATB-II targeting task for different machine learning models and using directed transfer function (DTF) as the feature. The table also shows the best set of hyperparameters during model calibration.

| Machine learning model | Optimal hyperparameters | Train RMSE (average Cross-validation score) | Test RMSE (average Cross-validation score) |
|---|---|---|---|
| Ridge Regression | alpha = 0.00126, fit intercept = False, solver = sag | 5.13 | 6.10 |
| Decision-tree regressor | min samples split = 4, min samples leaf = 8, max depth = 2 | 5.16 | 5.22 |
| Random forest regressor | num of estimators = 35, min samples split = 5, min samples leaf = 4, max depth = 9 | 5.18 | 5.26 |
| Support vector regressor | kernel = linear, gamma = 0.001, C = 1 | 5.41 | 5.78 |
| XGBoost regressor | subsample = 0.8, num of estimators = 150, max depth = 3, learning rate = 0.1, | 5.31 | 5.27 |
| | colsample bytree = 0.6 | | |
| Multi-layer perceptron | solver = adam, hidden layer sizes = (50,), alpha = 0.1, activation = ReLu | **5.11** | **4.97** |

**Table 3.** Train and test RMSEs for the average root mean square tracking error in the NASA-MATB-II targeting task for different machine learning models and using phase lag index (PLI) as the feature. The table also shows the best set of hyperparameters during model calibration.

| Machine learning model | Optimal hyperparameters | Train RMSE (average Cross-validation score) | Test RMSE (average Cross-validation score) |
|---|---|---|---|
| Ridge Regression | alpha = 3.19e-05, fit intercept = False, solver = sag | 5.13 | 6.31 |
| Decision-tree regressor | min samples split = 4, min samples leaf = 3, max depth = 8 | 6.83 | 7.93 |
| Random forest regressor | num of estimators = 20, min samples split = 2, min samples leaf = 3, max depth = 4 | 5.70 | 5.92 |
| Support vector regressor | kernel = linear, gamma = 0.001, C = 1 | 5.31 | 6.14 |
| XGBoost regressor | subsample = 0.6, num of estimators = 100, max depth = 3, learning rate = 0.1, column sample bytree = 1.0 | 5.85 | 6.14 |
| Multi-layer perceptron | solver = sgd, hidden layer sizes = (50,), alpha = 0.01, activation = tanh | **5.11** | **5.98** |

In addition, we also generated Shapley Additive Explanations (SHAP) graphs [17] to interpret the contribution and importance of each input feature on the model prediction. The SHAP graphs were designed for the best model in each of the functional connectivity/machine learning model combinations predicting both the output variables. Figure 2 shows the SHAP graph for the prediction of the average root mean square tracking error in the NASA-MATB-II task using a combination of multi-layer perceptron and the directed transfer function.

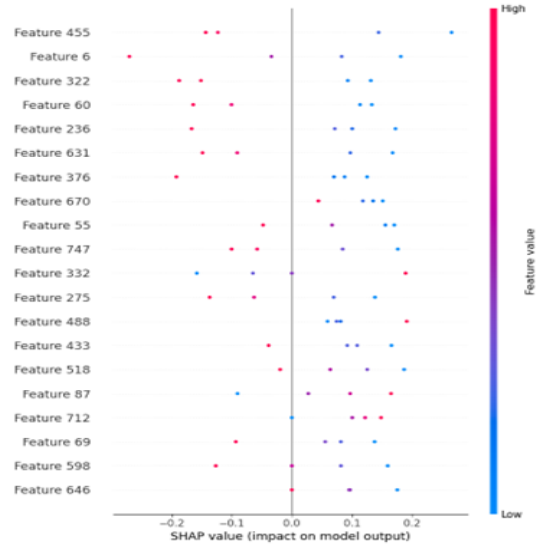

**Fig. 2.** SHAP graph for the average root mean square tracking error in the NASA-MATB-II task for the multi-layer perceptron using directed transfer function as the input feature.

As shown in Figure 2, the SHAP graph provides an insight into how each feature affects the output of the model for a specific instance or observation [17]. The y-axis of the SHAP graph represents the features in the dataset, while the x-axis represents the Shapley values associated with those features. Each feature is represented by a bar, and the length of the bar indicates the magnitude of the Shapley value. The color of the bar represents the value of the corresponding feature, where red indicates a high feature value and blue indicates a low feature value. Positive Shapley values indicate that the corresponding feature increases the prediction, while negative values suggest a decrease in the prediction. The longer the bar, the greater the contribution of that feature to the prediction. The SHAP graphs revealed a few functional connections between different ROIs identified as discriminative features in the prediction of cognitive performance. The functional connection between Brodmann Area 17 (the primary visual cortex, located in the occipital lobe) and Brodmann Area 24 (the anterior cingulate cortex, located in the cerebral cortex) [denoted as feature 455 in the SHAP graph) was shown to be the most discriminative feature in model prediction. Besides the individual feature contributions, a SHAP graph also provides an overview of the most important features and their impact on the model's predictions. It ranks the features based on their absolute Shapley values, allowing you to quickly identify the influential features.

## 5     Discussion and conclusion

This research aimed to efficiently predict cognitive performance in a complex multi-tasking paradigm using functional connectivity algorithms and machine learning models. Functional connectivity algorithms like directed transfer function (DTF) (based on Granger causality), phase locking value, and machine learning models like ridge regression, support vector regression, decision tree regression, random forest regression, XGBoost regression, and multi-layer perceptron were used. Results revealed that the DTF and multi-layer perceptron, yielded the lowest RMSE at around 5.11 for training and 4.97 for testing. Compared to prior EEG studies [4-6], we employed a source-estimation (dSPM) technique to correlate the recorded scalp activity to the appropriate brain activity. Source estimation methods have been shown to recover activated regions in the brain with high spatial accuracy [10]. The results obtained were consistent with [12], who stated that DTF's ability to estimate the direction of information flow between brain regions could help provide critical insights into the causal relationship between brain regions. In addition, according to [12], DTF has the propensity to capture temporal dynamics from the EEG signal over time, thus facilitating the successful interpretation of transient and state-dependent connectivity patterns. In addition, DTF has also been shown to be relatively robust to the effects of volume conduction compared to other source-space-based brain connectivity measures. This enables DTF to identify key network hubs and influential nodes in brain networks contributing to information processing in complex environments. These characteristics of DTF were exploited by the multi-layer perceptron, which is inherently capable of approximating complex non-linear relationships between the input and output variables [12]. In addition, the multi-layer perceptron's ability to learn hierarchical representations and represent higher-level features was useful in capturing relevant data patterns.

   Another vital contribution of this research work was the generation of SHAP graphs to improve the interpretability of the best-performing models to derive critical insights on model prediction eventually. The results from the SHAP graph were consistent revealed that the primary visual cortex has a central role in visual information processing, and the anterior cingulate cortex has been shown to have a fundamental role in attention allocation. These results were consistent with [4], where multitasking performance improved post-longitudinal tDCS administration, presumably due to the increased functional connections between the primary visual cortex and the anterior cingulate cortex, especially with the dorsal part of the anterior cingulate cortex connected to the prefrontal cortex, a part of the brain which is crucial for decision-making and higher-order cognitive processes. In addition, functional connections between the primary somatosensory cortex (Brodmann area 1) [denoted as feature 6], the orbitofrontal cortex (Brodmann area 12) [denoted as feature 322] were also found to be discriminative, indicating a higher number of functional connections created in the prefrontal cortex post-longitudinal tDCS, henceforth inducing significant changes in cognitive performance.

   There are some limitations in the current research work. A standard T1 image was employed during the source estimation procedure for all participants. Even though we used a cortical parcellation approach with relatively wide cortical areas to minimize estimation error, future studies utilizing the anatomy of individuals from high-resolution fMRI images may be able to increase source estimation accuracy. In addition, deep neural network architectures might be able to better understand and decipher the underlying brain dynamics contributing to cognitive performance post-longitudinal tDCS administration. In the future, the current machine learning algorithms can be improved to support multiclass prediction for individual tasks and across multiple tasks, which could result in more generalizable predictions for cognitive performance.

**References**

1. Van Orden, G. C., Holden, J. G., & Turvey, M. T. (2003). Self-organization of cognitive performance. Journal of experimental psychology: General, 132(3), 331.
2. Courage, M. L., Bakhtiar, A., Fitzpatrick, C., Kenny, S., & Brandeau, K. (2015). Growing up multitasking: The costs and benefits for cognitive development. Developmental Review, 35, 5-41.
3. Cammisuli, D. M., Cignoni, F., Ceravolo, R., Bonuccelli, U., & Castelnuovo, G. (2022). Transcranial Direct Current Stimulation (tDCS) as a Useful Rehabilitation Strategy to Improve Cognition in Patients With Alzheimer's Disease and Parkinson's Disease: An Updated Systematic Review of Randomized Controlled Trials. Frontiers in Neurology, 12, 2648.
4. Gupta, A., Daniel, R., Rao, A., Roy, P. P., Chandra, S., & Kim, B. G. (2023). Raw Electroencephalogram-Based Cognitive Workload Classification Using Directed and Nondirected Functional Connectivity Analysis and Deep Learning. Big Data.
5. Daniel, R., Pandey, V., Bhat, K. R., Rao, A. K., Singh, R., & Chandra, S. (2018, September). An empirical evaluation of short-term memory retention using different high-density EEG based brain connectivity measures. In 2018 26th European Signal Processing Conference (EUSIPCO) (pp. 1387-1391). IEEE.
6. Mahajan, R., Daniel, R. V., Rao, A. K., Pandey, V., Chauhan, R. P., & Chandra, S. (2021). Effect of Beta-Frequency Binaural Beats on Cognitive Control in Healthy Adults. In Proceedings of 6th International Conference on Recent Trends in Computing: ICRTC 2020 (pp. 685-698). Springer Singapore.
7. Karthikeyan, R., Smoot, M. R., & Mehta, R. K. (2021). Anodal tDCS augments and preserves working memory beyond time-on-task deficits. Scientific reports, 11(1), 19134.
8. Chase, H. W., Boudewyn, M. A., Carter, C. S., & Phillips, M. L. (2020). Transcranial direct current stimulation: a roadmap for research, from mechanism of action to clinical implementation. Molecular psychiatry, 25(2), 397-407.
9. Phang, C. R., Ting, C. M., Noman, F., & Ombao, H. (2019). Classification of EEG-based brain connectivity networks in schizophrenia using a multi-domain connectome convolutional neural network. arXiv preprint arXiv:1903.08858.
10. Tadel, F., Baillet, S., Mosher, J. C., Pantazis, D., & Leahy, R. M. (2011). Brainstorm: a user-friendly application for MEG/EEG analysis. Computational intelligence and neuroscience, 2011, 1-13.
11. Rasser, P. E., Johnston, P. J., Ward, P. B., & Thompson, P. M. (2004, April). A deformable Brodmann area atlas. In 2004 2nd IEEE International Symposium on Biomedical Imaging: Nano to Macro (IEEE Cat No. 04EX821) (pp. 400-403). IEEE.

12. Cao, J., Zhao, Y., Shan, X., Wei, H. L., Guo, Y., Chen, L., ... & Sarrigiannis, P. G. (2022). Brain functional and effective connectivity based on electroencephalography recordings: A review. Human Brain Mapping, 43(2), 860-879.
13. Arashi, M., Roozbeh, M., Hamzah, N. A., & Gasparini, M. (2021). Ridge regression and its applications in genetic studies. Plos one, 16(4), e0245376.
14. Yang, L., & Shami, A. (2020). On hyperparameter optimization of machine learning algorithms: Theory and practice. Neurocomputing, 415, 295-316.
15. Qiu, Y., Zhou, J., Khandelwal, M., Yang, H., Yang, P., & Li, C. (2021). Performance evaluation of hybrid WOA-XGBoost, GWO-XGBoost and BO-XGBoost models to predict blast-induced ground vibration. Engineering with Computers, 1-18.
16. Moayedi, H., Osouli, A., Nguyen, H., & Rashid, A. S. A. (2021). A novel Harris hawks' optimization and k-fold cross-validation predicting slope stability. Engineering with Computers, 37, 369-379.
17. Wang, J., Wiens, J., & Lundberg, S. (2021, March). Shapley flow: A graph-based approach to interpreting model predictions. In International Conference on Artificial Intelligence and Statistics (pp. 721-729). PMLR.